\documentclass[sigconf,9pt]{acmart}

\renewcommand\footnotetextcopyrightpermission[1]{} 
\AtBeginDocument{%
  \providecommand\BibTeX{{%
    \normalfont B\kern-0.5em{\scshape i\kern-0.25em b}\kern-0.8em\TeX}}}
\usepackage{booktabs} 
\usepackage[colorinlistoftodos,prependcaption,textsize=tiny]{todonotes}
\usepackage{enumitem}
\usepackage[nodisplayskipstretch]{setspace}

\usepackage{subcaption}

\graphicspath{{images/}}

\copyrightyear{2019} 
\acmYear{2019} 
\setcopyright{acmcopyright}
\acmConference[SCOPES '19]{22nd International Workshop on Software and Compilers for Embedded Systems}{May 27--28, 2019}{Sankt Goar, Germany}
\acmBooktitle{22nd International Workshop on Software and Compilers for Embedded Systems (SCOPES '19), May 27--28, 2019, Sankt Goar, Germany}
\acmPrice{15.00}
\acmDOI{10.1145/3323439.3323988}
\acmISBN{978-1-4503-6762-2/19/05}

\begin{document}

%
\title{Memory and Parallelism Analysis Using a Platform-Independent Approach}

%

\author{Stefano Corda$^{1, 2}$, Gagandeep Singh$^{1, 2}$, Ahsan Javed Awan$^3$, Roel Jordans$^1$, Henk Corporaal$^1$}
\affiliation{\vspace{-0.4cm} \normalsize $^1$Eindhoven University of Technology   \hspace{1cm}$^2$IBM Research - Zurich \hspace{1cm} $^3$Ericsson Research}
\email{{s.corda, g.singh, r.jordans, h.corporaal}@tue.nl, ahsan.javed.awan@ericsson.com}

%
\renewcommand{\shortauthors}{Corda, et al.}

\begin{abstract}

Emerging computing architectures such as near-memory computing (NMC) promise improved performance for applications by reducing the data movement between CPU and memory. However, detecting such applications is not a trivial task. In this ongoing work, we extend the state-of-the-art platform- independent software analysis tool with NMC related metrics such as memory entropy, spatial locality, data-level, and basic-block-level parallelism. These metrics help to identify the applications more suitable for NMC architectures.


\end{abstract}

%
%




\raggedbottom
\settopmatter{printfolios=false}
\settopmatter{printacmref=false}


\maketitle

\section{Introduction}
\label{sec:introduction}
With the demise of Dennard scaling and slowing of Moore's law, computing performance is hitting a plateau \cite{Esmaeilzadeh:2011:DSE:2024723.2000108}. Furthermore, the improvements in memory and processor technology have grown at different speeds, which is infamously termed as the memory wall ~\cite{Wulf:1995:HMW:216585.216588}.
Additionally, the current big-data era, where data is being generated in a massive amount and across multiple domains, has created a demand for novel memory-centric designs rather than conventional compute-centric designs \cite{overviewpaper}.

Therefore, it has been made even more crucial for computer designer understand the characteristics of these emerging applications to optimize future systems for their target workloads. Among the different approaches that have been used in the past for application characterization, a micro-architecture independent approach provides more relevant workload characteristics than by using e.g. HW performance counters.
In this scope, the platform-independent software analysis tool PISA \cite{Anghel2015AnIA} was developed. PISA is capable of extracting results in a true micro-architecture agnostic manner, by utilizing the LLVM compiler framework Intermediate Representation (IR). Therefore, we extend the capabilities of PISA to extract NMC related characteristics.




The rest of the paper is organized as follows: \emph{Section \ref{sec:backgrounds}} presents the background information concerning the tool and the related works.
In \emph{Section \ref{sec:metrics}} we describe the characterization metrics we embedded into PISA.
In \emph{Section \ref{sec:results}} we show and discuss the characterization results. 
Finally, \emph{Section \ref{sec:conclusions}} concludes this paper.


\section{Background and Related Work}
\label{sec:backgrounds}


PISA is based on the LLVM Compiler framework.
It uses an intermediate representation (IR), which is generated from the application source using a clang front-end, to represent the application code in a generic way. This IR is independent of the target architecture and has RISC-like instruction set. Therefore, these features can be used to perform application analysis or optimization using the opt tool.
LLVM's IR has a hierarchical structure: a \emph{basic-block} that consists of \emph{instructions} and represents a single entry and single exit section of code; a \emph{function} that is a set of basic-blocks; and a \emph{module} that represents the application and contains functions and global variables.


\begin{figure}[H]
\centering
\includegraphics[width=8.5cm]{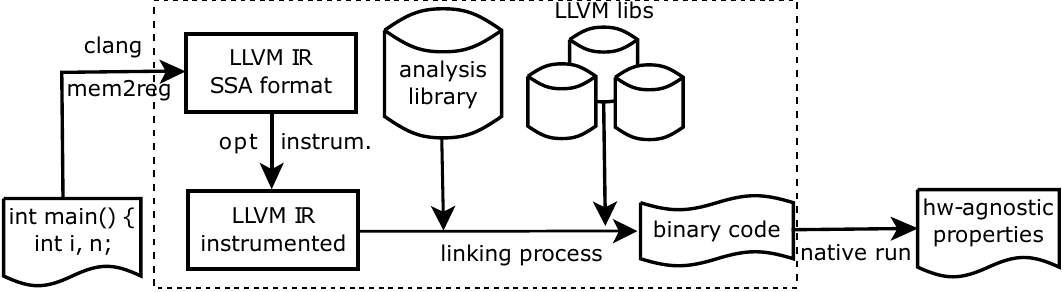}
\caption{Overview of the Platform-Independent Software Analysis Tool \cite{Anghel2015AnIA}.
\label{fig:pisa}}
\end{figure}

PISA's architecture is shown in \emph{Figure \ref{fig:pisa}}. Initially, the application source code, e.g. C/C++ code, is translated into the LLVM's IR. 
PISA exploits the \emph{opt} tool to perform LLVM's IR optimizations and to perform the instrumentation process using an LLVM pass. This process is done by inserting calls to the external analysis library throughout the application's IR.
The last step consists of a linking process that generates a native executable. On running this executable, we can obtain analysis results for specified metrics in JSON format. PISA can extract metrics such as instruction mix, branch entropy, data reuse distance, etc.

The analysis reconstructs and analyzes the program's instruction flow. This is possible because the analysis library is aware of the single entry and exit point for each basic-block. All the instructions contained in the basic-block are analyzed using the external library methods. 
Moreover, PISA supports the MPI and OpenMP standards allowing the analysis of multi-threaded and multi-process applications.
The tool's overhead depends on the analysis performed. On average the execution-time increases by two to three orders of magnitude in comparison to the non-instrumented code. However, since the analysis is target-independent, this has to be performed only once per application and dataset.

Considerable effort has been already spent in realizing platform independent characterization tools.
Cabezas \cite{cabezas2012tool} proposed a tool that can extract different features from workloads but has many limitations: the compiler community no longer supports the LLVM interpreter, and the target applications should be single threaded.
Another tool has been developed by Shao et al. \cite{shao}. It can extract interesting metrics such as memory entropy and branch entropy. However, this tool has some limitations: it is based on the IDJIT IR (just-in-time compilation) that has compatibility problems with OpenMP and MPI, thus being limited to sequential applications.
The state-of-the-art tool (called PISA) in workload characterization was presented by Anghel et al. \cite{Anghel2015AnIA}. PISA can analyze multi-threaded applications supporting the OpenMP and the MPI standards. PISA can extract the metrics such as instruction mix, branch entropy, data reuse distance, etc.
We extended PISA with metrics directed towards NMC such as memory entropy and spatial locality, data-level and basic-block-level parallelism. 

\section{Characterization Metrics}
\label{sec:metrics}

In this section we present the metrics we integrated into PISA. We 
focus on the memory behaviour, which is essential to decide if an application should be accelerated with a NMC architecture, and on the parallelism behaviour, which is crucial to decide if a specific parallel architecture should be integrated into an NMC system.




\subsection{Memory entropy}
The first metric related to memory behavior that we added is the memory entropy. The memory entropy measures the randomness of the memory addresses accessed. If the memory entropy is high, which means a higher cache miss ratio, the application may benefit from 3D-stacked memory because of the volume of data moved from the main memory to the caches. In information theory, Shannon's formula \cite{6773263} is used to capture entropy.

%


We embed in PISA, the formula defined by Yen et al. \cite{Yen:2008:NHT:1521747.1521793}. They applied Shannon's definition to memory addresses: $\textrm{Memory\_entropy} = - \sum^{2^n}_{i=1}\hat{p}(x_i)log_2\hat{p}(x_i)$, where $x_i$ is a n-bit random variable, $\hat{p}(x_i)$ is the occurrence probability for the value $x_i$ and $2^n$ is the number of values that $x_i$ can take. $\hat{p}(x_i)$ is defined by: $\hat{p}(x_i)=\frac{1}{d}\sum^{d}_{j=1}I(a_j=x_i)$ where $ I(a_j=x_i)=1 \; if \; (a_j=x_i), \; 0 \; \textrm{otherwise} \; \textrm{and} \; 0log0=0$.


In the last formula the addresses are represented as  $\{a_j\}^d_{j=1}$, where $d$ is the number of different addresses accessed during the execution. Each address is in the range $[0,2^{n-1}]$, where $n$ is the length of the address in bits. If every address has the same occurrence probability the entropy is $n$; if only one address is accessed the entropy is $0$. Otherwise the entropy is within $0$ and $n$. The memory entropy metric does not distinguish whether the accesses contain sequential patterns or random accesses. Therefore we need additional metrics, like spatial locality.

\subsection{Data reuse distance for multiple cache-line size and spatial locality}
Data reuse distance or data temporal reuse (DTR) is a helpful metric to detect cache inefficiencies. The DTR of an address is the number of unique addresses accessed since the last reference of the requested data. 
This metric is present in the default framework. However, the tool could compute it only for a fixed cache line size, which represents the address granularity. We extend the DTR computation and compute it starting from the word size to the value selected by the user. This extends the available analysis opportunities e.g. we use it to compute the spatial locality metric. 


Spatial locality, which measures the probability of accessing nearby memory locations, can be derived from DTR. We extend PISA with the spatial locality score inspired by Gu et al. \cite{Gu:2009:CMS:1542431.1542446}. The key idea behind this spatial locality score is to detect a reduction in DTR when doubling the cache line size. To estimate the spatial locality in a program two elements are fundamental: 1) histograms of data reuse distance for different cache line sizes, 2) distribution maps to keep track of changes in DTR for each access doubling the cache line size.
Histograms are used to compute the DTR distribution probability for different cache-line sizes. In \cite{Gu:2009:CMS:1542431.1542446} the reuse signature has been defined as a pair $<R,P>$, where $R$ is a series of consecutive DTR ranges of bins, represented as: $r_i=[d_i,d_{i+1})$. These bins are a logarithmic progression defined as:  $d_{i+1} = 2d_i (i \geq 0)$. $P$ is the distribution probabilities $p_i$ of the bin $r_i$. This reuse signature is used later to normalize the results.

The next step consists of building a distribution map. This map keeps track of each change in the DTR for every access. The distribution map has $i$ rows representing the bins using a cache line size $b$ and $j$ columns representing the bins using a doubled cache line size $2b$. Each cell is the probability $p_{ij}$ of the bin $i$ using a cache line size $b$ to change in a bin $j$ using a cache line size $2b$.  Differently from \cite{Gu:2009:CMS:1542431.1542446} we compute the sum of the cells in a row where $i<j.\;$
We do that because we want to express all the changes in data reuse distance. The spatial locality score for the bin $i$ is: $SLQ(i)=\sum_{j=0}^{j<i}p_{ij}$.


To compute the spatial locality score related to a pair of cache line sizes $<b,2b>$ we first compute the absolute values of the weighted sum that uses the probabilities $p_i$ included in the reuse signature and then use the formula proposed by \cite{Gu:2009:CMS:1542431.1542446} to calculate the total score, which is the logarithmic weighted sum of absolute values: $SLQ=\frac{\sum_{\textrm{all} \; b}|\sum_{\textrm{all} \; i}SLQ^b(i)p_i^b|2^{-b}}{\sum_{\textrm{all} \; b}2^{-b}}$.


The weighted score gives more importance to lower cache line sizes pairs.
Nevertheless, this can be interpreted as higher relevance of these lower pairs because bigger cache line sizes bring massive data transfers.
Usually, application with low spatial locality perform very bad on traditional systems with cache hierarchies because a small portion of data is utilized compared to the data loaded from the main memory to the caches.

\subsection{Data-level parallelism}
Data-level parallelism (DLP) measures the average length of vector instructions that is used to optimize a program. DLP could be interesting for NMC when employing specific SIMD processing units in the logic layer of the 3D-stacked memory.

\begin{figure}[H]
\centering
\includegraphics[width=7.8cm,trim ={0 0.5cm 0 1.8cm},clip]{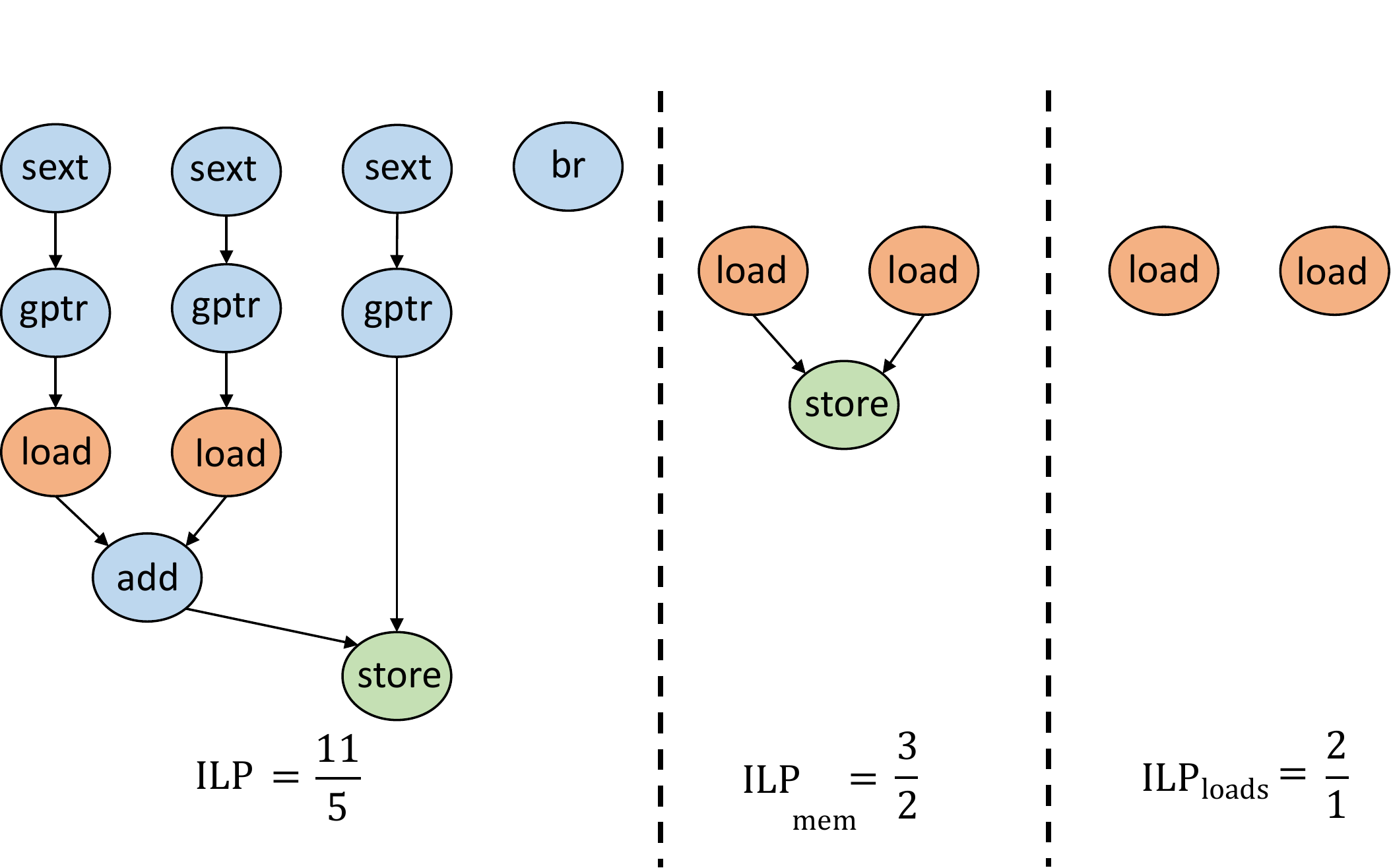}

\caption{
Usually to compute ILP it has been used the control flow graph (CFG), left side.
The CFG in the center is used to compute the ILP per type of instructions. On the right our per opcode specialized CFG.\label{fig:ilpspecialized}}
\end{figure}

PISA can extract the instruction-level parallelism for all the instructions (see \emph{Figure \ref{fig:ilpspecialized}}, CFG on the left) and additionally per instruction category such as control, memory, etc. (see \emph{Figure \ref{fig:ilpspecialized}}, CFG in the center). As shown in the CFG on the right in \emph{Figure \ref{fig:ilpspecialized}}, we extract the ILP score per opcode and call it as $ ILP_{\textrm{specialized}, \textrm{opcode}}$ where opcode can be load, store, add, etc. This metric represents the number of instructions with the same opcode that could run in parallel. Next, we compute the weighted average value for DLP using the weighted sum over all opcodes of $ ILP_{\textrm{specialized}, \textrm{opcode}}$. The weights are the frequency of the opcodes calculated by dividing the number of instructions per code with the number of instructions.

$DLP_{avg}=\sum_{\textrm{opcode}} ILP_{\textrm{specialized}, \textrm{opcode}} \frac{\#\textrm{instructions}_{\textrm{opcode}}}{\#\textrm{instructions}}$ 


As the register allocation step is not performed at the level of intermediate representation, it is not possible to take into account the register consecutiveness in this score.
However, we want to show the optimization opportunities for compilers distinguishing between consecutiveness of load/store instruction addresses. We represent this with two scores: $DLP_1$ without address consecutiveness; $DLP_2$ with addresses consecutiveness into account. To compute them we use the previous formula changing the $ILP_{\textrm{specialized}, \textrm{opcode}}$ value for loads and stores.

\subsection{Basic-block level parallelism}

A basic-block is the smallest component in the LLVM's IR that can be considered as a potential parallelizable task. Basic-block level parallelism (BBLP) is a potential metric for NMC because it can estimate the task level parallelism in the application. The parallel tasks can be offloaded to multiple compute units located on the logic layer of a 3D-stacked memory.

\begin{figure}[H]
\centering
\includegraphics[width=8.5cm,trim ={0 10.3cm 12.9cm 3.1cm},clip]{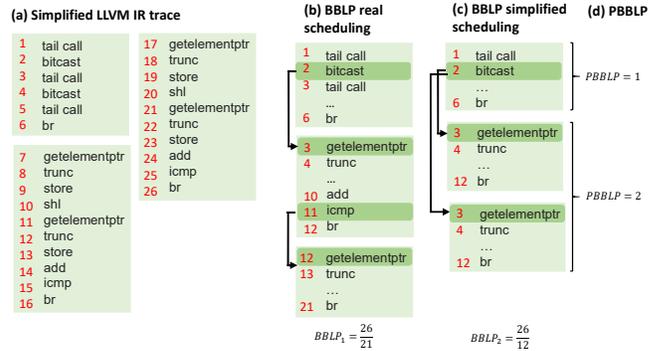}
\caption{BBLP/PBBLP methodology: a) example of LLVM dynamic trace; b) real scheduling for BBLP computation taking in account all dependencies; c) simplified scheduling for BBLP computation not taking in account dependencies such as loop index update (in a) dependency between instruction 15 and 17); d) PBBLP values for each basic block (second and third block are a repeated basic block, since there is only a loop index dependecy, the PBBLP is equal to 2). 
\label{fig:bblpscheduling}}
\end{figure}

To estimate BBLP in a workload, we develop a metric similar to ILP and DLP. It is based on the assumption that  a basic-block, which is a set of instructions, can only be executed sequentially. Since loop index count could put an artificially tight constraint on the parallelism, we assume two different basic-block scheduling approaches (see \emph{Figure \ref{fig:bblpscheduling}}): 1) all the dependencies between basic-block are considered; 2) we consider a smart scheduling, assuming a compiler that can optimize loop index update dependencies. The difference between the two approaches can give an idea, as in the DLP case, of the optimization opportunities for compilers. We compute the two scores derived from the two scheduling options using the following formula: $BBLP_{avg}=\frac{\# \textrm{instructions}}{\textrm{MaxIssueCycle}_{BBLP}}$, where $MaxIssueCycle_{BBLP}$ represents the cycle of the last executed instruction using the proposed scheduling approaches (red numbers in \emph{Figure \ref{fig:bblpscheduling}(b,c)}). $\#instructions$ represent the total number of instructions (see \emph{Figure \ref{fig:bblpscheduling}.a}). 

We also aim to estimate the presence of data parallel loops. Data parallel loops consists of basic-blocks that are repeated without any dependencies among their instances. A fast and straightforward estimation can be done by assigning a value to each basic-block between $1$ and the number of instances. When a basic-block has only one instance or all its instances have dependencies among them the score is $1$. Instead, when all its instances don't have dependencies among them the value is maximal and equal to the number of instances. Contrariwise, the score is within the range described above.
Other assumptions we made are: skip index update dependencies and omit basic-blocks that are used only for index update.

After assigning a score to each basic-block ($PBBLP_{BB}$),  we compute the weighted average value for PBBLP using the weighted sum over all scores ($PBBLP_{BB}$). The weights are the frequency of the basic-block instances calculated by dividing the number of instances per basic-block with the number of total instances. $PBBLP_{avg}=\sum_{BB}PBBLP_{BB}\frac{\#\textrm{instances}_{BB}}{\#\textrm{instances}_{\textrm{total}}}$. Since this metric is an estimation we call it as potential basic-block level parallelism (PBBLP).

 

\section{Characterization results}
\label{sec:results}

\begin{figure}[H]
\centering

\includegraphics[width=8.5cm,trim={3cm 1cm 3cm 1cm},clip]{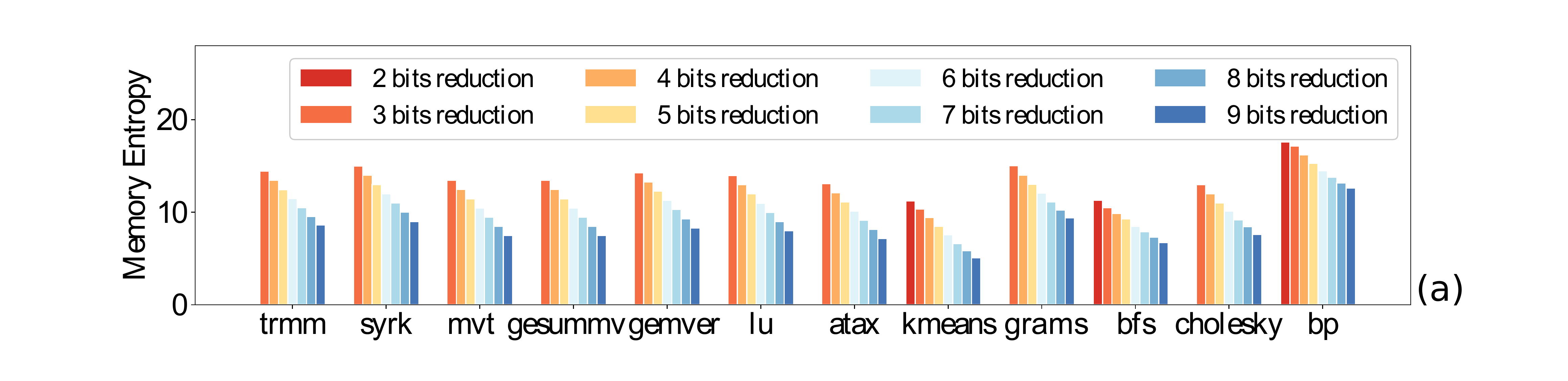}
\includegraphics[width=8.5cm,trim={3cm 1cm 3cm 1cm},clip]{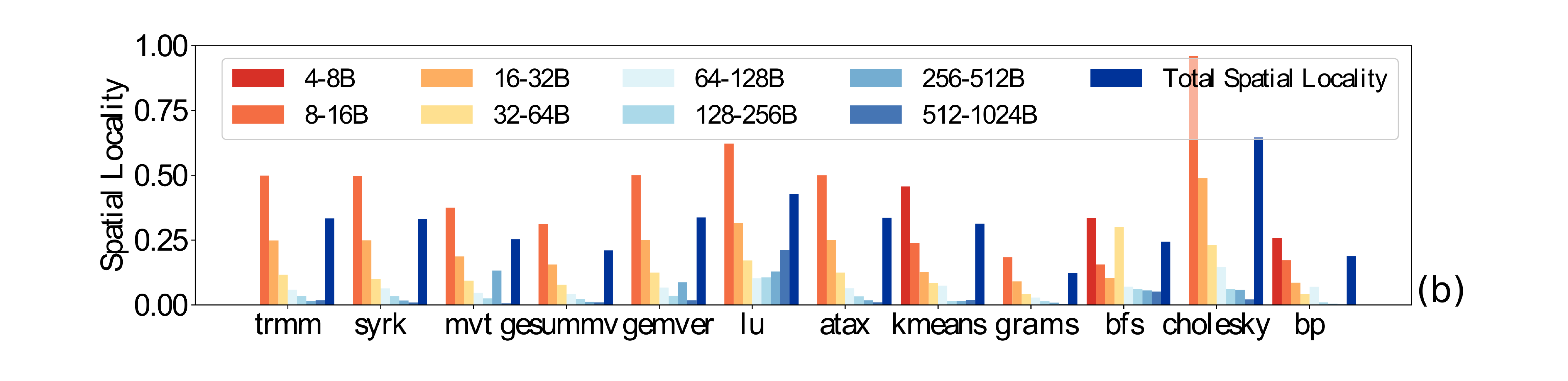}
\includegraphics[width=8.5cm,trim={3cm 1cm 3cm 1cm},clip]{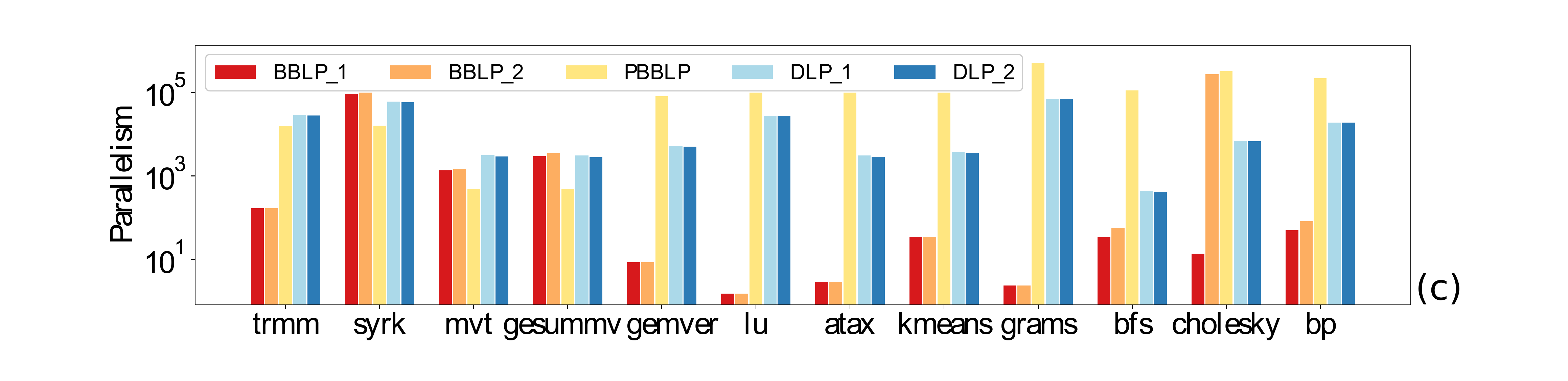}

\caption{Application characterization results:(a) Memory Entropy; (b) Spatial Locality; (c) Parallelism.} \label{fig:analysis}

\end{figure}

We present the 
the characterization results of selected applications from PolyBench~\cite{pouchet2012polybench} and Rodinia~\cite{5306797} benchmarks (see \emph{Figure \ref{fig:analysis}}) employing the proposed metrics. 
Memory entropy, in \emph{Figure \ref{fig:analysis}.a}, is strictly related to the dimension of the address space accessed by a workload. Indeed, applications with larger address space have higher entropy because they are accessing many different addresses. We also plot memory entropy changes at different granularity cutting the least-significant bits (LSBs) of the address to represent larger data access granularity. Furthermore, we highlight in Rodinia's applications the cut of 2 LSBs because they are accessing integer (4Byte locations). We notice that applications like \texttt{bp} and \texttt{gramschmidt} have higher values of entropy and they should benefit from NMC architectures. Contrariwise, the other applications have similar values except for \texttt{cholesky}, \texttt{bfs} and \texttt{kmeans}. 

Related to memory behavior, we show in \emph{Figure \ref{fig:analysis}.b} the spatial locality of the workloads. As expected, we can distinguish different behaviors among the benchmarks. \texttt{bp} and \texttt{gramschmidth} show an interesting behavior with high entropy and low spatial locality. For instance, in \texttt{gramschmidt} accesses to the matrix are done by column and diagonally. However, the matrix allocation is done in a row-major order. These applications should be good candidates for NMC because they use a large address space with low locality. An opposite trend is detected for \texttt{cholesky}, where the entropy is one of the lowest value and the spatial locality is the highest value.

A considerable amount of applications show a spatial locality lower than 0.25 and they should benefit from NMC systems. 
However, applications with high spatial locality like \texttt{cholesky} could also benefit from NMC mostly when increasing the data-set and consequently moving more data off-chip and exploiting SIMD architectures.

\emph{Figure \ref{fig:analysis}.c} shows the parallelism characterization of workloads. As expected in the Berkeley dwarfs for the data-level parallelism analysis~\cite{asanovic2006landscape}, matrix multiplication based algorithms show the highest values.  Moreover, the difference between the two proposed DLP scores seems to be very limited. Only small variations can be noticed, for instance in \texttt{trmm} and \texttt{syrk}. Here, the difference is due to loads/stores with non-sequential accesses and could be improved by a compiler exploiting data mapping techniques. Instead the BBLP scores show a significant difference for \texttt{cholesky} and limited differences for \texttt{bfs} and \texttt{syrk}. These results highlight possible parallelism optimizations that can be performed by compilers.

Finally, the $PBBLP$ score tries to highlight the presence of data parallel loops and gives an estimation of how much parallelism can be achieved using vectorization or loop unrolling strategies.
Applications with high level of parallelism could benefit from NMC systems that provided multicores or SIMD architectures in the logic layer on top of the 3D-stacked memory.

\section{Conclusions}
\label{sec:conclusions}
Emerging computing architectures in their first stages of development such as near-memory computing (NMC) lack proper tools for specialized workload profiling.
In this scope, we have extended PISA, a state-of-the-art application characterization tool, with NMC related metrics. Particularly, we have concentrated on analyzing the memory accesses and parallelism behaviors: data-level parallelism, basic-block level parallelism, memory entropy, and spatial locality.
In a separate work we will explain the correlation between the proposed metrics and the performance on an NMC system.



  \begin{acks}
This work was performed in the framework of Horizon 2020 program and is funded by European Commission under Marie Sklodow- ska-Curie Innovative Training Networks European Industrial Doctorate (Project ID: 676240). We would like to thank Fetahi Wuhib and Wolfgang John from Ericsson Research for their feedback on the draft of the paper.
\end{acks}

\bibliographystyle{ACM-Reference-Format}
\bibliography{acmart}

\end{document}